\documentclass[aps,pra,preprint,showkeys,groupedaddress,showpacs,nofootinbib]{revtex4-1}
\usepackage{graphicx, subfigure}
\usepackage{amsthm,amssymb,amsmath,amsfonts,bm}

\DeclareMathOperator{\sign}{ sgn}
\begin{document}
	
	\title{Minlos-Faddeev regularization of zero-range interactions in the three-body problem}
	
	\author{O.~I.~Kartavtsev}\email{oik@nusun.jinr.ru}
	\author{A.~V.~Malykh}\email{maw@theor.jinr.ru}
	\affiliation{Joint Institute for Nuclear Research, 6 Joliot-Curie St, Dubna, Moscow Region, Russia, 141980} 
	

\begin{abstract} 
To regularize the three-body problem, Minlos and Faddeev suggested 
a modification of zero-range model, which diminishes interaction 
at the triple-collision point. 
The analysis reveals that this regularization results in four alternatives 
depending on the regularization parameter $ \sigma $. 
Explicitly, Efimov or Thomas effects remain for $ \sigma < \sigma_c $, 
the additional boundary conditions of two types should be imposed at 
the triple-collision point for $ \sigma_c \le \sigma < \sigma_e $ and 
$ \sigma_e < \sigma < \sigma_r $, and the problem is regularized 
for $ \sigma \ge \sigma_r $. 
Critical values $ \sigma_c < \sigma_e < \sigma_r $ separating different 
alternatives are determined both for a two-component three-body system 
and for three identical bosons. 
\end{abstract}


\maketitle

In studies of the universal low-energy dynamics, e.~g., in ultra-cold 
quantum gases, it is natural to use a zero-range model for short-range 
two-body interactions. 
Nevertheless, it is not a trivial task to introduce this in the few-body 
problem, which requires additional efforts. 
A basic origin of complication is connected with overlap of at least two 
zero-range interactions, which takes place near the triple-collision 
point~\cite{Minlos61}. 

Minlos and Faddeev proposed a specific modification of zero-range model 
in the three-body problem~\cite{Minlos61}. 
Twenty years later an equivalent form written as a boundary condition 
in the configuration space was given in~\cite{Albeverio81}. 
Both papers showed that an additional three-body force of the strength 
$ \sigma $ is able to preclude the Efimov and Thomas effects for sufficiently 
large $ \sigma $ exceeding the explicitly given critical value $ \sigma_c $. 
Moreover, the condition $ \sigma \ge \sigma_c $ was assumed to provide 
an unambiguous description of the three-body problem. 
Recently, a proof of this conjecture was presented for three identical 
bosons~\cite{Basti21} and for $ N $ identical bosons interacting with a 
distinct particle~\cite{Ferretti22}. 

The Minlos-Faddeev regularization of zero-range model is studied in this 
note both for the two-component system consisting of two identical bosons 
interacting with a distinct particle and for the system containing three 
identical bosons. 
The analysis is based on reducing the problem to consideration of  
an ordinary differential equation with singular coefficients. 
It is shown that the proposed regularization leads to different results 
in four intervals of the parameter $ 0 < \sigma < \infty $ separated by three 
critical values $ \sigma_c $, $ \sigma_e $, and $ \sigma_r $. 

More exactly, in the interval $ \sigma_c \le \sigma < \sigma_r $, it is 
necessary to set an additional boundary condition in the triple-collision 
point, which depends on an arbitrary real-valued parameter $ b $. 
Moreover, in the interval $ \sigma_e \le \sigma < \sigma_r $, this boundary 
condition should be written in a special form by taking account of the lower order 
terms in expansion near the triple-collision point. 
To elucidate consequences of the described regularization, dependence 
of the bound-state energy for three identical bosons on $ b $ and $ \sigma $ 
is determined for $ \sigma_c \le \sigma < \sigma_r $. 


Consider the system consisting of a distinct particle $ 1 $ of mass $ m_1 $ 
and two identical bosons $ 2 $ and $ 3 $ of masses $ m_2 = m_3 = m $. 
In the center-of-mass frame, define the scaled Jacobi variables as 
$ {\mathbf x} = \displaystyle\sqrt{2\mu}\left({\mathbf r}_2 - 
{\mathbf r}_1\right) $ and 
$ {\mathbf y} = \displaystyle\sqrt{2\tilde\mu}\left({\mathbf r}_3 - 
\dfrac{m_1{\mathbf r}_1 + m{\mathbf r}_2}{m_1 + m} \right) $, where 
$ {\mathbf r}_i $ is a position vector of $ i $th particle and 
the reduced masses are denoted by $ \mu = \dfrac{mm_1}{m + m_1} $ and 
$ \tilde{\mu} = \dfrac{m(m + m_1)}{m_1 + 2m} $. 
In the zero-range model, the two-body interaction is determined by a single 
parameter, the scattering length $ a $. 
The units are chosen as $ \hbar = |a| = 2 \mu = 1 $, which gives the unit 
two-body binding energy ($ \epsilon_{2b} = 1 $). 
The mass ratio $ m/m_1 $ remains a single essential parameter that 
for convenience can be interchangeably replaced by the kinematic angle 
$ \omega $ defined by $ \sin \omega = 1/(1 + m_1/m) $. 

The Hamiltonian is formally defined as the six-dimensional Laplace operator 
supplemented by the boundary conditions imposed at zero distance between 
the interacting particles. 
In paper~\cite{Minlos61} it was suggested to introduce in the momentum-space 
integral equation a term containing the convolution-type operator $ K(k - k') $ 
depending on the relative momentum $ k $ between the interacting pair's 
center-of-mass and the third particle, whose asymptotic form 
$  K(\xi ) \to \dfrac{\tilde\sigma}{\xi^2} $ for $ \xi \to \infty $. 
The equivalent form of this regularization in the configuration-space 
representation was proposed in paper~\cite{Albeverio81} and considered recently 
in~\cite{Basti21,Ferretti22}, where an additional term was introduced 
to modify the boundary conditions at zero distance between the interacting 
particles, 
\begin{equation}
\label{bound1}
\lim_{x \rightarrow 0}
\left[ \dfrac{\partial \log (x \Psi )}{\partial x} 
-\dfrac{\sigma }{\cos\omega }\frac{\theta(y)}{y} \right] = -\sign (a) \, .
\end{equation} 
Here $ \sigma $ is a regularization parameter, which controls the wave 
function near the triple-collision point, the factor 
$ \dfrac{1}{\cos\omega } $ is introduced for convenience, and $ \theta (y) $ 
is an arbitrary bounded function normalized by $ \theta (0) = 1 $ and 
$ \dfrac{d \theta }{d y}\bigg|_{y = 0} = 0 $. 
Only one boundary condition of the form~(\ref{bound1}) should be imposed if 
the symmetry under permutations of identical particles is taken into account. 

Clearly, the described procedure adds a kind of the three-body interaction, 
thus providing the regularization of Hamiltonian in the triple-collision point. 
As is well established, the zero total angular momentum and positive 
parity $ P $ ($ L^P = 0^+ $) are those quantum numbers, for which 
the regularization is certainly required, therefore, namely this case will 
be considered in this note. 
One should mention that the regularization is also needed for even $ L \ge 2 $ 
and positive parity if the mass ratio $ m/m_1 $ of the two-component system is 
sufficiently large~\cite{Kartavtsev07a,Endo11,Helfrich11,Kartavtsev19}. 
Nonetheless, description of the regularized problem for $ L \ge 2 $ is 
analogous to the case of $ L = 0 $ and will not be given in this work. 


In study of the proposed regularization, a key point is to determine 
its impact on the wave function in the triple-collision point. 
To proceed, one introduces the hyper-radius $ \rho $ and hyper-angular 
variables on a hyper-sphere  
$ \{\Omega \} = \{ \alpha, \hat{\mathbf x}, \hat{\mathbf y} \} $, which are 
defined by $ x = \rho\cos\alpha $, $ y = \rho \sin \alpha $ and 
$ \hat{\mathbf x} = {\mathbf x}/x$, and $\hat{\mathbf y} = {\mathbf y}/y $. 
As in~\cite{Kartavtsev07,Kartavtsev07a,Kartavtsev16,Kartavtsev19}, 
the total wave function is expanded in a set of functions 
$ \Phi_n(\alpha, \hat{\mathbf x}, \hat{\mathbf y}; \rho) $, which are 
the solutions of an auxiliary eigenvalue problem on a hyper-sphere 
(for fixed $ \rho $), 
\begin{equation}
\label{Psi}
\displaystyle
\Psi ({\mathbf x}, {\mathbf y}) = \rho^{-5/2} \sum_{n = 1}^{\infty} f_n(\rho) 
\Phi_n(\alpha, \hat{\mathbf x}, \hat{\mathbf y}; \rho)\, . 
\end{equation}

The functions $ \Phi_n(\alpha, \hat{\mathbf x}, \hat{\mathbf y}; \rho) $ 
for $ L^P = 0^+ $ satisfy the equation 
\begin{equation}
\label{eqonhypershere}
\hskip -.2cm
\left[\dfrac{\partial^2}{\partial\alpha^2} + 4 \tan 2\alpha 
\dfrac{\partial}{\partial\alpha} + \gamma^2 (\rho ) 
- 4 \right] \Phi(\alpha, \hat{\mathbf x}, \hat{\mathbf y}; \rho) = 0 \ , 
\end{equation} 
the boundary condition  
\begin{equation}
\label{bch}
\lim_{\alpha \rightarrow 0} \frac{\partial \log 
	\left[ \alpha \Phi(\alpha, \hat{\mathbf x}, \hat{\mathbf y};  \rho) \right] }
{\partial \alpha }  = -\rho \, \sign (a) + \dfrac{\sigma \ \theta(\rho)}{\cos\omega } 
 \end{equation} 
and inherit the permutation symmetry of the total wave function. 
Solving the eigenvalue problem~(\ref{eqonhypershere}) and~(\ref{bch}), 
one obtains the equation 
\begin{equation} 
\label{transeql0}
\left[\rho \sign(a) - \frac{\sigma \ \theta(\rho )}{\cos \omega }  \right] 
\sin \gamma \frac{\pi }{2} = \gamma \cos \gamma \frac{\pi }{2} - 
2 \frac{\sin \gamma \omega}{\sin 2 \omega } 
\end{equation} 
for the two-component system. 
Similarly, for three identical bosons, taking into account that 
$ \omega = \pi/6 $ and all three particles interact, one finds 
\begin{equation} 
\label{transeql03id}
\hskip -.38cm 
\left[\rho \sign(a) - \dfrac{2 \sigma \, \theta(\rho )}{\sqrt{3}}  \right] 
\sin \gamma \dfrac{\pi }{2} = \gamma \cos \gamma \dfrac{\pi }{2} - 
\dfrac{8}{\sqrt{3}} \sin \gamma \dfrac{\pi }{6}. 
\end{equation} 
The eigenvalue equations~(\ref{transeql0}) and (\ref{transeql03id}), 
whose different branches determine an infinite set of eigenvalues 
$ \gamma^2_n(\rho ) $, are of the well-known 
form~\cite{Kartavtsev99,Kartavtsev07a,Kartavtsev19}, in which
the regularization gives rise to an additional term proportional to 
$ \sigma \ \theta(\rho ) $ \footnote{The eigenvalue equations, which are 
formally written for a finite two-body scattering length $ a $, remain 
applicable both for $ a \to \infty $ (unitary two-body interaction) and 
for $ a = 0 $ (non-interacting particles) by taking $ \rho = 0 $ 
for the former and $ \rho \to \infty $ for the latter.}. 

By means of the expansion~(\ref{Psi}), the original problem is formulated as 
a set of coupled hyper-radial 
equations~\cite{Kartavtsev07,Kartavtsev16,Kartavtsev19} for the channel 
functions $ f_n(\rho) $, 
\begin{eqnarray}
\left[\frac{d^2}{d \rho^2} - \frac{\gamma_n^2 - 1/4}{\rho^2} + 
P_{nn} + E \right] f_n(\rho ) -  
\sum_{m \neq n}^{\infty}\left[P_{nm} - 
Q_{nm} \frac{d}{d\rho} - \frac{d}{d\rho}Q_{nm} \right] f_m(\rho ) = 0 \ , 
\label{system1} 
\end{eqnarray}
where matrix elements
$ Q_{nm}(\rho ) = \displaystyle 
\int  \Phi_n \dfrac{\partial\Phi_m}{\partial\rho} d \Omega$ and 
$ P_{nm}(\rho) = \displaystyle \int 
\dfrac{\partial\Phi_n}{\partial\rho}\dfrac{\partial\Phi_m}{\partial\rho} d \Omega $ 
can be given in the analytical form via $ \gamma_n(\rho ) $ and their 
derivatives~\cite{Kartavtsev99,Kartavtsev06,Kartavtsev07}. 


The diagonal terms in the hyper-radial equations~(\ref{system1}), which 
are singular near the triple-collision point ($ \rho \to 0 $), are of principal  
importance for analysis of the three-body problem and its regularization. 
The matrix elements $ Q_{nm}(\rho ) $ and $ P_{nm}(\rho) $, 
as follows from their analytical 
expressions~\cite{Kartavtsev99,Kartavtsev06,Kartavtsev07}, 
are finite and differentiable functions for any $ \rho $ and do not influence 
essential properties of the solution near $ \rho = 0 $. 
Explicitly, the principal singularity comes from the branch corresponding to
the smallest $ \gamma_1^2(\rho ) $, whose singular part is
\begin{equation}
\label{V1_zero}
V_\mathrm{sing}(\rho ) = \dfrac{\tilde{\gamma }^2 - 1/4}{\rho^2} + \dfrac{q}{\rho}  
\end{equation} 
for $ \rho \to 0 $. 
Here the notations $ \tilde{\gamma } \equiv \gamma_1(0) $ 
and $ q \equiv \displaystyle \frac{d\gamma_1^2}
{d\rho } \bigg|_{\rho = 0} $ are introduced for brevity. 
A number of aspects concerning the Schr{\"o}dinger equation with inverse square 
singularity, as in $ V_\mathrm{sing}(\rho ) $, were multiply discussed 
in literature, e.~g., in~\cite{Case50,Braaten04,Bouaziz14,Kartavtsev19}. 
Following~\cite{Kartavtsev16,Kartavtsev19}, the essential features of 
the quantum problem for the singular potential of the form 
$ V_\mathrm{sing}(\rho ) $~(\ref{V1_zero}) are briefly summarized below. 

There are four variants for unambiguous description of the problem, which 
correspond to four intervals of the real-valued $ \tilde{\gamma }^2 $, as 
is understood by considering two solutions $ f_{\pm} (\rho) $ of 
the Schr{\"o}dinger equation with the potential $ V_\mathrm{sing}(\rho ) $, 
whose leading-order terms for $ \rho \to 0 $ are given by 
$ f_{\pm} (\rho) \sim \rho^{ 1/2 \pm \tilde{\gamma } } $. 
Consider first the case $ \tilde{\gamma}^2 \ge 1 $, for which one of 
the solutions, $ f_-(\rho ) $, is not square integrable and should be ruled out, 
thus, the problem is defined by the condition of square integrability or 
simply by $ f(\rho) \xrightarrow[\rho \to 0]{} 0 $. 
On the other hand, for $ \tilde{\gamma }^2 < 1 $ both solutions 
$ f_{\pm}(\rho ) $ are square integrable for $ \rho \to 0 $ and any linear 
combination of them is acceptable. 
To get rid of this ambiguity, it is necessary to choose a specific linear 
combination by introducing one additional real-valued parameter. 
In particular, for $ \tilde{\gamma }^2 < 0 $, i.~e., pure imaginary 
$ \tilde{\gamma } $, both solutions 
$ f_\pm (\rho) \sim \rho^{1/2 \pm i |\tilde{\gamma }| } $ oscillate for 
$ \rho \to 0 $. 
One possibility to define the unambiguous problem in this case is to fix 
a relative phase $ \delta $ of two oscillating functions, i.~e., by 
the requirement $ f(\rho) \xrightarrow[\rho \to 0]{} \rho^{1/2} 
\sin (|\tilde{\gamma}| \log (\rho) + \delta ) $~\cite{Case50,Braaten04}. 
As an important consequence, one obtains the asymptotic energy spectrum, which 
depends exponentially on the level's number, 
$ E_n \sim e^{- 2 \pi n/|\tilde{\gamma}|} $. 
In fact, these considerations explain the Efimov effect in the three-body 
problem~\cite{Efimov73,Fedorov93,Kartavtsev07a}. 
One should also notice that the asymptotic energy spectrum of this form was 
derived already in~\cite{Minlos61}. 

For the interval $ 1/4 \le \tilde{\gamma}^2 < 1 $, it is necessary to take into 
account the less singular term $ q/\rho $ in 
$ V_\mathrm{sing}(\rho ) $~(\ref{V1_zero}) because the next-to-leading 
order term in $ f_-(\rho) \sim  \rho^{1/2 - \tilde{\gamma}} 
\left[ 1 + q \rho /(1 - 2 \tilde{\gamma} ) \right] $ is of principal importance. 
Thus, to define the problem for $ 0 \le \tilde{\gamma}^2 < 1 $ one should 
introduce an additional real-valued parameter $ b $ by imposing the boundary 
condition, e.~g., of the form proposed in~\cite{Kartavtsev16,Kartavtsev19} 
\begin{equation}
\label{as_gam121}
 f(\rho ) \xrightarrow[\rho \to 0]{ } \rho^{\frac{1}{2} + \tilde{\gamma}} 
- \sign (b) |b|^{2 \tilde{\gamma}} \rho^{\frac{1}{2} - \tilde{\gamma}} 
\left(1 + \dfrac{q \rho}{1 - 2 \tilde{\gamma} }\right), 
\end{equation} 
where the $ q $-dependent term can be omitted for 
$ 0 \le \tilde{\gamma}^2 < 1/4 $. 

Special treatment is needed for two critical values $ \tilde{\gamma}^2 = 0 $ 
and $ 1/4 $ separating different types of the problem definition. 
In particular, for $ \tilde{\gamma}^2 = 0 $ the boundary condition could 
be written in the form 
\begin{equation}
\label{as_gam0}
f( \rho) \xrightarrow[ \rho \to 0]{}  \rho^{\frac{1}{2}} \, \log  \dfrac{\rho}{ b} \ , 
\end{equation} 
where only positive values of the parameter are admitted ($ b > 0 $). 
For $ \tilde{\gamma}^2 = 1/4 $, one of the solution is 
$ f_- \sim 1 + q \rho \log \rho $ for $ \rho \to 0 $ and the $ q $-dependent 
boundary condition can be written in the form, 
\begin{equation}
\label{as_gam12}
f(\rho) \xrightarrow[ \rho \to 0]{}  \rho - b (1 + q  \rho \log  \rho) \ . 
\end{equation}

\begin{figure*}[!t] 
\centering
\includegraphics[width=0.48\textwidth]{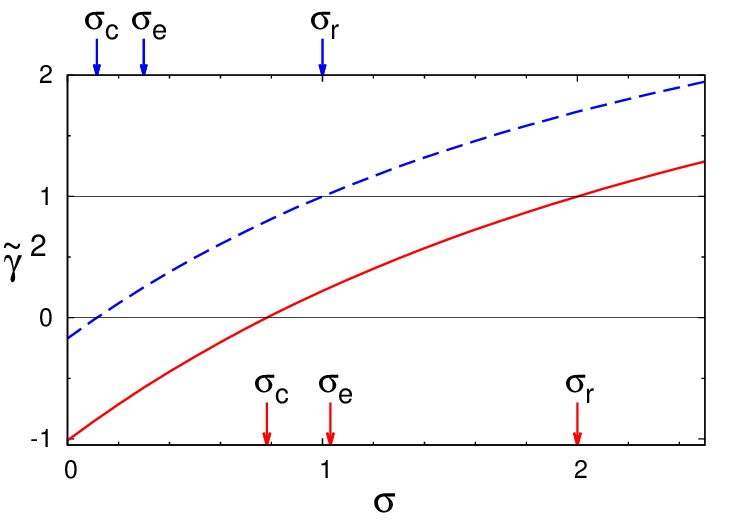}
\includegraphics[width=0.48\textwidth]{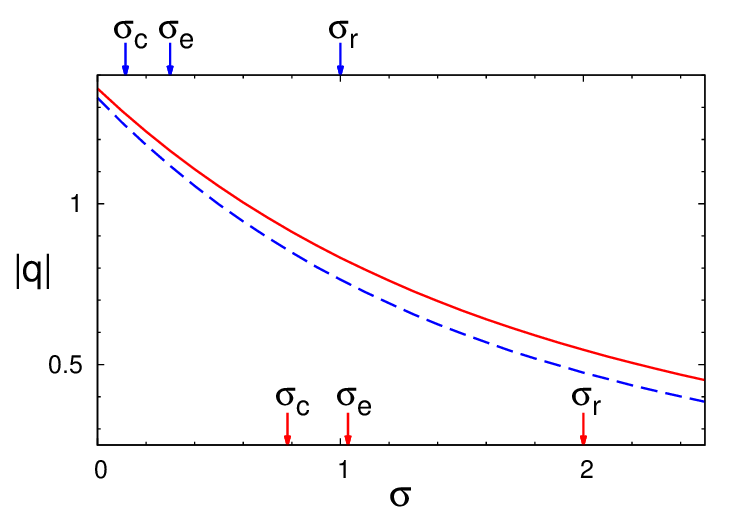}
\caption{ 	{\bf Fig. 1.} 
Dependencies $ \tilde{\gamma}^2 $ (left) and $ |q| $ (right) on 
the regularization parameter $ \sigma $ are plotted by solid (red) lines for 
three identical bosons and by dashed (blue) lines for two identical bosons 
and a distinct particle of the same mass ($ m/m_1 = 1 $). 
Arrows indicate the critical values $ \sigma_c $, $ \sigma_e $, and 
$ \sigma_r $ at lower (upper) border for the former (latter) case. 
\label{fig_gammaq}  
}  
\end{figure*}

The main result of this letter is a complete analysis of the Minlos-Faddeev 
regularization by using the explicitly given dependencies $ \tilde{\gamma } $ 
and $ q $ on $ \sigma $. 
Taking $ \rho = 0 $ in~(\ref{transeql0}) and~(\ref{transeql03id}), one finds 
\begin{equation}
\label{sigma}
\sigma =  \frac{\dfrac{\sin \tilde{\gamma } \omega }{\sin \omega } - 
	\tilde{\gamma } \cos \omega \cos \tilde{\gamma } \dfrac{\pi }{2}}
{\sin \tilde{\gamma } \dfrac{\pi }{2}}
\end{equation} 
for two identical bosons and a distinct particles and
\begin{equation}
\label{sigma3id}
\sigma =  \frac{4 \sin \tilde{\gamma } \dfrac{\pi }{6} - 
	\dfrac{\sqrt{3}}{2} \tilde{\gamma } \cos \tilde{\gamma } \dfrac{\pi }{2}}
{\sin \tilde{\gamma } \dfrac{\pi }{2}}
\end{equation} 
for three identical bosons. 
Both Eqs.~(\ref{sigma}) and~(\ref{sigma3id}) implicitly determine monotonically 
increasing functions $ \tilde{\gamma }^2 (\sigma) $, which are shown 
in Fig.~\ref{fig_gammaq} both for the two-component system in the case 
$ m/m_1 = 1 $ and for three identical bosons. 
Without regularization, i.~e., if $ \sigma = 0 $, the well-known values of pure 
imaginary $ \tilde{\gamma } \approx i \, 0.4136973 $ for the two-component 
system with $ m/m_1 = 1 $ and $ \tilde{\gamma } \approx i \, 1.0062378 $ for 
three identical bosons are obtained. 
Another parameter entering in the boundary condition for $ \rho \to 0 $ 
is expressed via $ \tilde{\gamma } $ by using its definition and Eqs.~(\ref{transeql0}) and~(\ref{transeql03id}), which gives for two identical 
bosons and a distinct particles 
\begin{eqnarray}
\label{depq}
\vspace{.5cm} q =  
\dfrac{4 \sign(a)\tilde{\gamma}  
	\sin^2 \tilde{\gamma } \dfrac{\pi }{2}}
{\sin \tilde{\gamma } \pi - \tilde{\gamma } \pi 
- \dfrac{4\omega \cos \tilde{\gamma} \omega \sin \tilde{\gamma } \dfrac{\pi }{2} 
- 2\pi \sin \tilde{\gamma}\omega \cos \tilde{\gamma } \dfrac{\pi }{2}}{\sin 2 \omega} } 
\end{eqnarray} 
and for three identical bosons 
\begin{equation}
\label{depq3id}
q = \dfrac{ 4 \sign(a)\tilde{\gamma} \sin^2 \tilde{\gamma } \dfrac{\pi }{2}}
{\sin \tilde{\gamma }\pi -  \tilde{\gamma } \pi - \dfrac{32 \pi } 
	{3 \sqrt{3}} \sin \tilde{\gamma } \dfrac{\pi}{3} \sin^2 \tilde{\gamma } 
	\dfrac{\pi }{6}} \ . 
\end{equation} 
The dependencies $ q(\sigma ) $ for the two-component system in the case 
$ m/m_1 = 1 $ and for three identical bosons are depicted 
in Fig.~\ref{fig_gammaq}. 

As follows from the above results, the regularization suggested by 
Minlos and Faddeev gives rise to four separate outcomes depending on the value 
of $ \sigma $, which is a consequence of the correspondence with 
$ \tilde{\gamma }^2 $. 
Different types of the regularized problem are separated by the critical values 
$ \sigma_c $, $ \sigma_e $, and $ \sigma_r $, which correspond to 
$ \tilde{\gamma}^2 = 0 $, $ 1/4 $, and $ 1 $, respectively. 
Explicitly, the three-body problem becomes regularized for 
$ \sigma \ge \sigma_r $ and one real-valued parameter should be introduced 
to impose the boundary condition of the form~(\ref{as_gam121}) 
in the triple-collision point for $ \sigma_r > \sigma \ge \sigma_c $. 
Furthermore, the condition should be $ q $-dependent for 
$ \sigma_r > \sigma > \sigma_e $, whereas $ q $-dependence can be safely 
omitted for $ \sigma_e > \sigma \ge \sigma_c $. 
At last, the Efimov or Thomas effect takes place for $ \sigma_c > \sigma $ 
and the famous exponential dependence on the level's number, 
$ E_n \sim e^{- 2 \pi n/|\tilde{\gamma}|} $ could be obtained by introducing 
one-parameter boundary condition in the triple-collision point. 
At two critical points $ \sigma = \sigma_c $ and $ \sigma = \sigma_e $ 
one should use the specific boundary conditions~(\ref{as_gam0}) 
and~(\ref{as_gam12}), respectively. 

\begin{figure}[h]    
\centering     
\includegraphics[width=0.48\textwidth]{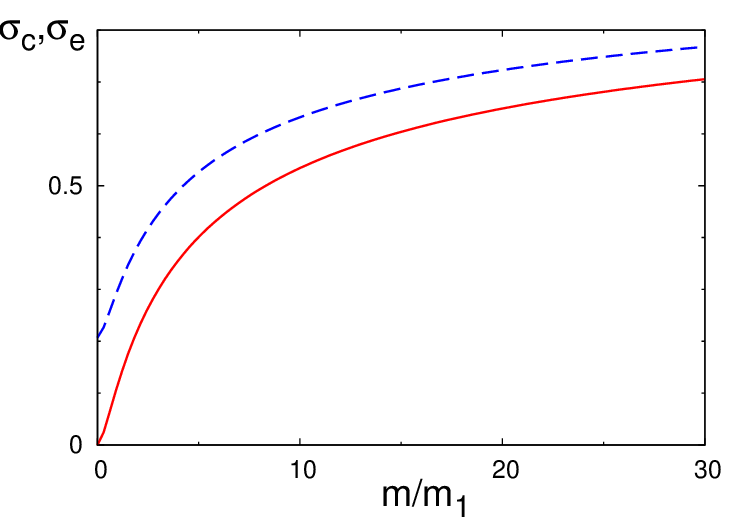}
\caption{	{\bf Fig. 2.} 
	Critical values $ \sigma_c $ (solid red line) and $ \sigma_e $ 
	(dashed blue line) as functions of the mass ratio $ m/m_1 $ for 
	the two-component system. }  
\label{fig_sigmam}
\end{figure}

As follows from~(\ref{sigma}), the critical values for the two-component system 
are given by 
\begin{eqnarray} 
\label{sigmac}
\sigma_c = \dfrac{2}{\pi} \left( \frac{\omega }{\sin \omega } - 
\cos \omega \right) \ , 
\end{eqnarray} 
\begin{eqnarray} 
\label{sigmae}
\sigma_e = \frac{1}{\sqrt{2}\cos \dfrac{\omega}{2}} - \frac{1}{2} \cos\omega \ , 
\end{eqnarray}
and $ \sigma_r = 1 $. 
The dependencies $ \sigma_c(m/m_1) $ and $ \sigma_e(m/m_1) $ are depicted 
in Fig.~\ref{fig_sigmam}. 
For the case of equal masses ($ m/m_1 = 1 $), from~(\ref{sigma}) one finds 
$ \sigma_r = 1 $, $ \sigma_e = 3 \sqrt{3}/4 - 1 \approx 0.29904$, and 
$ \sigma_c = 2/3 - \sqrt{3}/\pi  \approx 0.11534 $. 
This value of $ \sigma_c $ was also given in~\cite{Ferretti22}. 
Similarly, from~(\ref{sigma3id}) one finds the critical values $ \sigma_r = 2 $, 
$ \sigma_e = 7\sqrt{3}/4 - 2 \approx 1.03109 $, and 
$ \sigma_c = 4/3 - \sqrt{3}/\pi \approx 0.78200 $ for three identical bosons. 
Exactly this value of $ \sigma_c $ was also given 
in papers~\cite{Minlos61,Albeverio81,Basti21}. 

It is worthwhile to present three values of $ q $ 
corresponding to $ \sigma_c $,  $ \sigma_e $, and $ \sigma_r $, namely, 
$ |q(\sigma_c)| = 486/[ \pi (81 + 8 \sqrt{3}\pi )] \approx 1.24225 $, 
$ |q(\sigma_e)| = 18/(8\sqrt{3} \pi - 18 - 3 \pi ) \approx 1.11757 $, and 
$ |q(\sigma_r)| = 12/(5 \pi ) \approx 0.76394 $ from Eq.~(\ref{depq}) 
for the two-component system in the case $ m/m_1 = 1 $ and 
$ |q(\sigma_c)| = 486/[\pi (81 + 16 \sqrt{3}\pi )] \approx 0.920483 $, 
$ |q(\sigma_e)| = 18/[(16 \sqrt{3} - 15) \pi - 18] \approx 0.820476 $, and 
$ |q(\sigma_r)| = 12/(7 \pi ) \approx 0.545674 $ from Eq.~(\ref{depq3id}) 
for three identical bosons. 

\begin{figure}[!t] 
	\centering
	\includegraphics[width=0.47\textwidth]{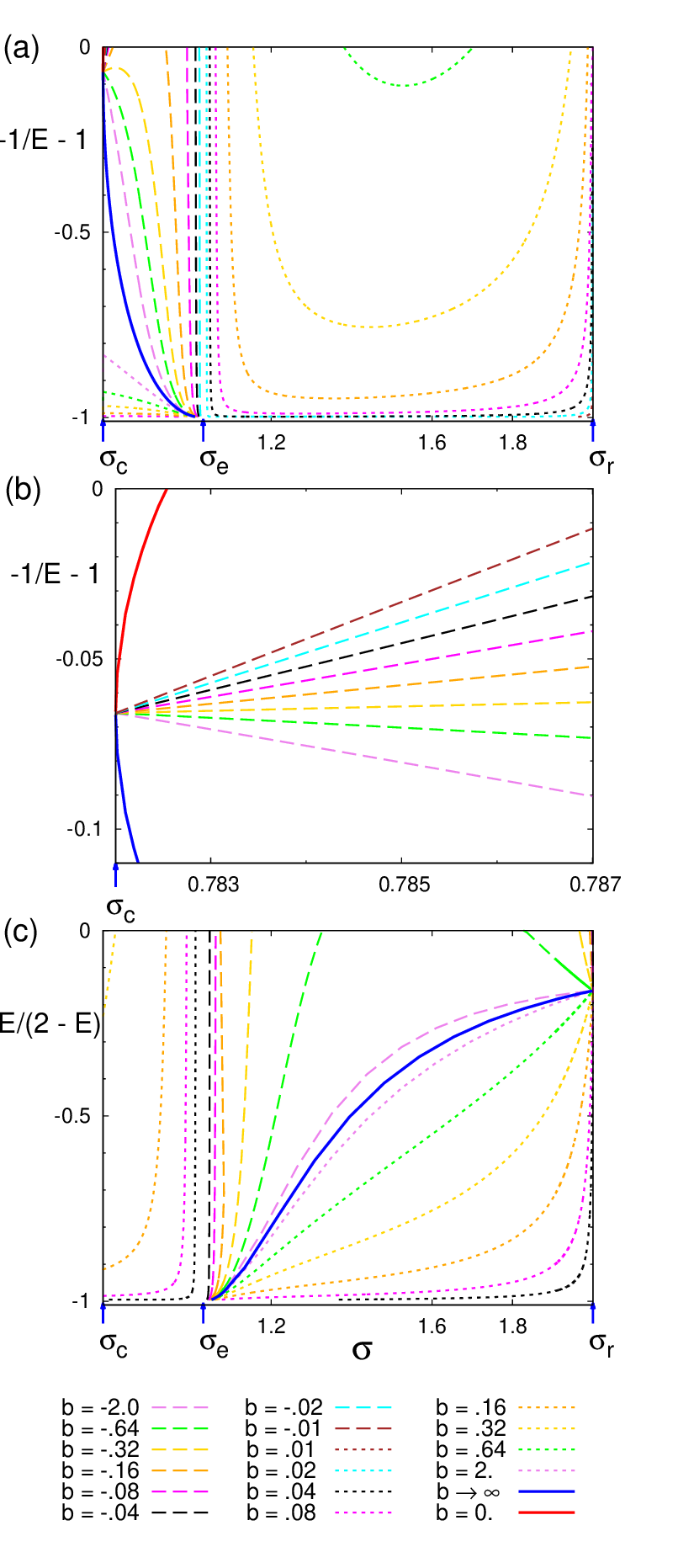}
	{\caption{
	{\bf Fig. 3.}
			The bound-state energy $ E $ of three identical bosons as a function of 
			the regularization parameter $ \sigma $ for different values of $ b $ 
			is depicted for the positive [panels (a) and (b)] and negative [panel (c)] 
			two-body scattering length. 
			A small area near $ \sigma_c $ in panel (a) is magnified and shown in panel (b). 
			To represent all below-threshold values of energy, the ordinate axes are scaled 
			to map either $ -\infty < E < -1 $ in panels (a) and (b) or $ -\infty < E < 0 $ 
			in panel (c) onto the interval $ (-1, 0) $. 
		} 
		\label{fig_en}}
	\vspace{-.5cm}
\end{figure}

To elucidate the above results, the bound-state energies of three identical 
bosons are determined as a function of $ \sigma $ and $ b $ in the interval 
$ \sigma_c \le \sigma < \sigma_r $. 
It is fulfilled by solving a truncated system of hyper-radial 
equations~(\ref{system1}) complemented by the boundary condition in 
the triple-collision point of the form~(\ref{as_gam121}) or~(\ref{as_gam0}) 
for $ \sigma_c $ and~(\ref{as_gam12}) for $ \sigma_e $. 
In this calculation, a simple function $ \theta(y) = 1 $ is chosen to specify 
the boundary condition~(\ref{bound1}). 
Up to eight hyper-radial equations has been used to reach a typical accuracy 
of about five - six significant. 
The calculated energy dependence is presented in Fig.~\ref{fig_en}. 

It turns out that at most one bound state exists and the bound-state energy 
monotonically increases with increasing $ b $ at fixed $ \sigma $.  
If $ b = 0 $, a bound state exists only for $ a > 0 $ and its energy 
monotonically increases from $ E_c \approx -1.0670 $ to the threshold 
$ E_\mathrm{th} = -1 $ with increasing $ \sigma $ within the small 
interval $ \sigma_c \le \sigma \le \sigma_0 \approx 0.7825 $. 
If $ b \to \infty $, there is a bound state, whose energy for $ a > 0 $ 
decreases from $ E_c $ to $ - \infty $ with $ \sigma $ increasing  from 
$ \sigma_c $ to $ \sigma_e $; and for $ a < 0 $ increases from $ - \infty $ 
to $ E_r \approx -0.38792 $ with $ \sigma $ increasing from $ \sigma_e $ 
to $ \sigma_r $. 
Only positive $ b $ are admitted at the critical value $ \sigma_c $, therefore, 
all the energies for $ b \le 0 $ tend to same value $ E_c $ for 
$ \sigma \to \sigma_c $. 
At another boundary of the interval, all the existing energies tend to $ E_r $ 
for $ a < 0 $ and disappear for $ a > 0 $. 
Recall that there are no bound states for $ \sigma \le \sigma_r $. 

In the limit $ |a| \to \infty $, the hyper-radial equations~(\ref{system1}) 
become decoupled and one bound state exists for $ b > 0 $, whose energy is 
$ E = -\dfrac{4}{b^2} \left[ - \dfrac{\Gamma (\tilde{\gamma })}
{\Gamma(-\tilde{\gamma })} \right]^{1/\tilde{\gamma }} $ and the channel 
function is $ f(\rho) = \rho^{1/2} K_{\tilde{\gamma } }(\sqrt{-E} \rho ) $, 
where $ \tilde{\gamma } $ is related to $ \sigma $ by Eq.~(\ref{sigma3id}) and 
$ K_\nu(x) $ is a modified Bessel function. 


Starting from the first suggestion to modify the two-body zero-range 
interaction proposed by Minlos and Faddeev~\cite{Minlos61} it was 
declared~\cite{Albeverio81,Basti21,Ferretti22} that the three-body problem 
becomes regularized, if the regularization parameter $ \sigma $ 
is sufficiently large to suppress the Efimov or Thomas effects, i.~e., if  
$ \sigma $ exceeds the critical value $ \sigma_c $ defined in~(\ref{sigmac}).  

In this work it was shown that the proposed regularization leads to different 
results in four intervals of the non-negative parameter $ \sigma $, 
in particular, another and more strict condition 
$ \sigma > \sigma_r > \sigma_c $ is necessary for unambiguous description 
of the three-body problem. 
Concerning the interval $ \sigma_c \le \sigma < \sigma_r $, it is necessary 
to set a boundary condition in the triple-collision point depending on one 
real-valued parameter $ b $. 
Among different possibilities, the boundary condition depending on 
the additional parameter $ b $ is chosen in this work as~(\ref{as_gam121}) 
for the interval $ \sigma_c < \sigma < \sigma_r $ and (\ref{as_gam0}) or  
(\ref{as_gam12}) for two specific values of $ \sigma $. 
It is essential that for the smaller interval 
$ \sigma_e \le \sigma < \sigma_r $, the boundary condition contains 
the parameter $ q $, which is exactly determined by Eqs.~(\ref{depq}) 
and~(\ref{depq3id}). 
At last, the Efimov or Thomas effects are present for $ \sigma < \sigma_c $ 
and the exponential asymptotics of the energy spectrum takes place after 
introducing the boundary condition in the triple-collision point. 

To exemplify in details the main conclusions, three critical values 
$ \sigma_c $, $ \sigma_e $, and $ \sigma_r $ are determined both for 
the two-component system consisting of two identical bosons and a distinct particle 
and for the system consisting of three identical bosons. 
The effect of regularization is additionally demonstrated by the calculation 
of the bound-state energy for three identical bosons as a function of
$ \sigma $ and $ b $. 

It is worthwhile to mention that the described scenario is quite general. 
In fact, it could be anticipated for any problem, 
whose essential properties are determined by the effective potential with 
the singular part $ \sim \rho^{-2} $, which strength goes through the critical 
values. 
Besides the two-component system consisting of two identical particles 
(either fermions or bosons) and a distinct one, which was described 
in~\cite{Kartavtsev16,Kartavtsev19}, this scenario could be of importance also 
for the three-body problem in the mixed 
dimensions~\cite{Nishida08a,Nishida11,Lamporesi10} or 
in the presence of spin-orbit interaction~\cite{Shi14,Cui14,Shi15}.

\end{document}